\documentclass[a4paper,12pt]{article}
\usepackage[pdftex]{graphicx}
\usepackage{caption}
\usepackage{amsmath}
\usepackage{plain}
\title{The distributions of violent event and interevent times in conflicts}

\author{Jeroen Bruggeman\thanks{Department of Sociology, University of Amsterdam, Nieuwe Achtergracht 166, 1018 WV Amsterdam, the Netherlands. Email: j.p.bruggeman@uva.nl. ORCID: 0000-0002-0422-1842}}
\date{\today }

\begin{document}
\maketitle






\begin{abstract}
Violent conflicts at all scales are interrupted by shorter and longer lulls without violence. Studies of interevent times found power-law distributions of interevent times, based on coarse-grained data with a resolution of days or longer. When interevent times are represented as multiplicative processes, however, a mathematical theorem predicts a lognormal distribution for fine-grained data, not a power law. Here, video-recorded street fights between groups of young men are used as fine-grained data. The behavior of individuals in these groups has been hand-coded on the timeline, pointing out when fighting takes place and is interrupted. It turns out that the lognormal does not fit better than the power law (and vice versa), and the outcome is inconclusive. Violent events, by contrast, take much energy and on average, their durations are shorter than interevents. Hence, their distribution has a short tail that is well-approximated by a lognormal distribution.





\end{abstract}

\section*{Introduction}
Violence is common across human history, and occurs between entities ranging from individuals to empires \cite{mann23}. Although fights may look chaotic, statistical patterns have been found across large numbers of violent conflicts \cite{johnson08}. A case in point is the number of deadly casualties, which is power-law distributed \cite{johnson13,richardson48,cirillo16,cederman11}. Furthermore, when fighting erupts, it is occasionally suspended due to haphazard troubles that confrontations into violent events and interevents. It turns out that also the number of casualties per violent event is power-law distributed \cite{scharpf14,restrepo20}, as well as the durations of the interevents \cite{bohorquez09,okamoto23,picoli14}.

This paper looks closer at the distribution of interevent times within violent confrontations. During small scale conflicts, for example between hooligans, interruptions occur because individuals stumble, opponents move out of the way, or bystanders and obstacles stand in the way. During larger scale conflicts between groups with weapons and other resources, the list of potential problems is longer: ammunition runs out, equipment fails, logistics stall, and so on. If both camps face problems at the same time, fighting is suspended. Then, the fighters try to get their act together before their opponents strike. The earlier studies that found power-law distributed interevent times \cite{bohorquez09,okamoto23,picoli14} were based on data with a resolution of at least one day. However, if the resolution is much more fine-grained, a theorem predicts that the distribution is lognormal \cite{sornette97}, not power law. This is interesting because it proposes that long lulls are rarer than has been taken for granted, and might be an artifact of the data. Accordingly, I investigate if the power law is refuted when using data with a resolution five orders of magnitude smaller than the earlier studies. My data set consists of video recorded street fights between small groups of young men. The outcome is inconclusive, though. I also investigate the distribution of the event times. Because fighting takes much more energy than not-fighting, the tail of the event distribution is shorter.

\section*{Theory}
Assume that in a clash between two groups (or individuals), the first violent event has just ended, and the participants prepare for the next. Preparatory actions take a certain amount of time, $\tau$, and in the longer run or over many fights, a distribution of $\tau$ settles down.   
Take the first interevent time $\tau_1$ as a baseline. The next interevent time, $\tau_2$, can be shorter or longer than $\tau_1$, expressed by $v_1$, such that $\tau_2 = v_1 \tau_1$. After a sequence of $t+1$ violent events, the next interevent time can be written as $\tau_{t+1} = v_t v_{t-1} \ldots v_1 \tau_1$, or $log(\tau_{t+1}) =  log(\tau_1) + \sum^{k=t}_{k = 1} v_k$, where all $v$'s are random and independent with finite mean and variance. Now, the sequence of interevent times is represented as a multiplicative process \cite{gibrat30,mitzenmacher04}. Multiplicative processes have been used to model distributions of social phenomena such as city sizes and word (e.g., surname) usages \cite{zanette07}, but not yet for conflict data. 

Sornette and Cont \cite{sornette97} have proved that in a multiplicative process, $\tau$ approximates a lognormal distribution, 
\begin{equation}
P(\tau > s) = \int_{\tau = 0}^{\infty} \frac{1}{\sqrt{2 \pi}\sigma \tau} e^{(ln(\tau) - \mu)^2/2 \sigma^2} d \tau,
\end{equation}
where $s$ indicates time in seconds (Figure~\ref{fig:cumulative_dist}), $\mu = \overline{\tau}$, and  $\sigma^2$ is the variance of $\tau$.
Yet, they have also proved that if the minimum duration is not arbitrarily close to zero, the distribution is a power law,
\begin{equation}
P(\tau > s) \sim \tau^{-\alpha},
\end{equation}
with exponent $\alpha$.
Power laws have been found in earlier studies of violence based on coarse-grained data. Following Sornette and Cont, I conjecture that my fine-grained interevent data is closer to a lognormal than a power law. I add an exponential distribution to the comparison, 
\begin{equation}
P(\tau > s) \sim e^{-\lambda \tau},
\end{equation}
although the theory used here \cite{sornette97} makes no prediction about it. It adds a potential falsifier to the hypothesis, which now becomes squeezed in between the longer power-law tail and the shorter exponential tail. 

For violent event times, the same multiplicative representation can be used. Because violence takes a great deal of energy, event times are shorter than interevent times, which has been noticed long ago \cite{olzak89}. 
Due to the high energy demand, the tail of the event distribution will be short, and given the two distributions under comparison \cite{sornette97}, I conjecture that the event distribution is closer to a lognormal than a power law, leaving open the possibility that it is exponential.  

\section*{Materials and Methods}
My data of violence are video-recorded street fights of 59 small groups ($2 \leq N < 10; \overline{N} = 3.6$) attacking other small groups or individuals. They were recorded by bystanders on mobile phones, who posted their clips on websites such as LiveLeak, YouTube, and WorldStarHipHop; they last in between half a minute and five minutes. The clips were collected for another study, of violence as a collective action problem \cite{bruggeman26a}, using the keywords ``street fight,'' ``brawl,''  and ``assault''. The interevent and event times are random even though the selected videos are not. Behavior was coded as violence when force was used against another’s body (stomping, slapping, kicking) or when another’s body was forcefully moved, and was plotted on the timeline using Noldus Observer XT 14 software. During violent events, at least one individual acted violently; during interevent times, nobody did. The data and its examination are detailed in \cite{bruggeman26a}; this information together with the code (see the links at the end of this paper) allows readers to replicate the results. 

\begin{figure}[!ht] 
\captionsetup{width=.905\linewidth}
\begin{center}  
\includegraphics[width=0.45\textwidth]{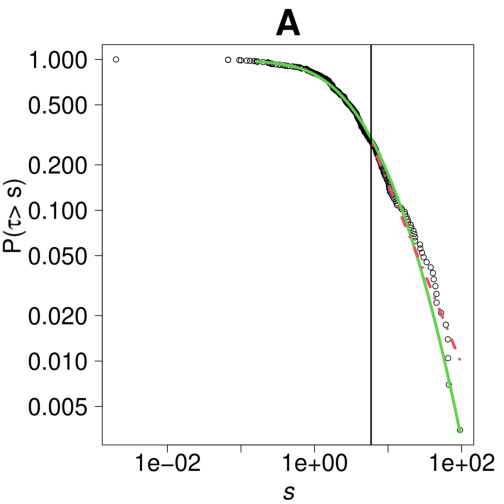}
\includegraphics[width=0.45\textwidth]{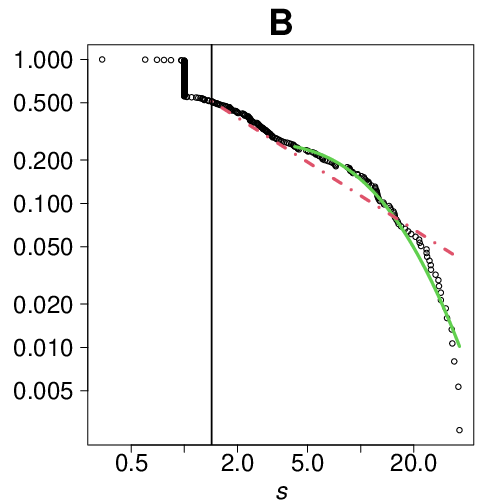}
\end{center} 
\caption{Power laws (continuous lines) and lognormals (dash-dotted) fitted to the data. The vertical lines indicate the cutoff points of the power laws. {\bf A}. Cumulative distribution, $P(\tau > s)$, of interevent times in seconds, $s$. {\bf B}. Cumulative distribution of event times. The jump is an artifact of the coding (see main text). }
\label{fig:cumulative_dist}
\end{figure}

For the cumulative distributions in Figure~\ref{fig:cumulative_dist}, I used maximum likelihood estimation of R's poweRlaw package, based on Clauset's method \cite{clauset09}. The lognormals have continuous lines and the power laws have dash-dotted lines; the traits of these distributions can be found in Table~\ref{table:stats}. 
\begin{table}[ht]
    \centering
    \begin{tabular}{ |l|l|l|l|l|l|l| } \hline
Data & Distribution  & $\tau_{min}$ & $n$  & Parameter(s) with sd  & KS    & $p$   \\ \hline 
int.ev. & power law & 5.88 & 83  & $\alpha = 2.2$ \, (0.175)        & 0.057  & 0.336 \\ 
int.ev. & lognormal  & 0.17 & 278 & $\mu = 1.14; \sigma = 1.53$   & 0.032  & 0.254 \\
int.ev. & exponential & 10.6 & 38 & $\lambda = 0.049$ \, (0.021)     & 0.0626 & 0.765  \\  \hline
event & power law    & 1.43 & 192 & $\alpha = 1.77$ \, (0.597 )     & 0.096  & 0  \\  
event & lognormal    & 4.23 & 93  & $\mu = 2.37; \sigma = 0.69$   & 0.057  & 0.107  \\ 
event & exponential & 3.17 & 108 & $\lambda = 0.103 $ \, (0.0217)   & 0.056  & 0.2993 \\ 
\hline 
\end{tabular}
 \caption{Distributions fitted to the data. The top three rows pertain to the interevent times, the bottom three to the event times. The columns present the cutoff value of the tail ($\tau_{min}$), the number of observations in the tail ($n$), the estimated parameters and their standard deviations, as well as the goodness of fit (Kolmogorov-Smirnov) and the $p$-values.}
\label{table:stats}
\end{table}

Over 59 fights there were 287 interevent times, ranging from 0.002 to 95 seconds.  
The goodness of fit (Kolmogorov-Smirnov) of the power law tells us that its fit to the data is plausible (KS $= 0.0574; p = 0.336$), but also the lognormal has a plausible fit (KS $= 0.032; p = 0.254$). The two distributions can be directly compared by means of Vuong's test, which yields a standardized log-likelihood ratio (Table~\ref{table:comparisons}). For this test, the tails' cutoffs are set to the same value, namely power-law's cutoff ($\tau_{min} = 5.88 s$, leaving $n = 83$). In this comparison, the lognormal is neither better nor worse than the power law (Table~\ref{table:comparisons}). Therefore, the first conjecture is not supported. However, the lognormal covers more data than the power law ($n = 278$ vs. 83), and therefore appears to be qualitatively better, although there is no statistical evidence to back this up. 
The exponential covers only 38 data points and cannot be taken seriously as a contender. When adding it to Figure~\ref{fig:cumulative_dist}A, it is hard to distinguish from the lognormal and because it clogs the image, it is not depicted. 

\begin{table}[ht]
    \centering
    \begin{tabular}{ |l|l|l|l|l|l| } \hline
Data      & Comparison               &  $\tau_{min}$ &  $n$ & Log-L & $p$     \\ \hline
int.ev    & lognormal and power law  &      5.88     &  83  &  0.837         & 0.201   \\
int.ev    & power law and lognormal  &      5.88     &  83  &  -0.837        & 0.799   \\
event     & lognormal and power law  &      4.23     &  93  &  3.46          & 0.00027 \\
event     & exponential and lognormal &     3.17     & 108  &  1.03          & 0.151 \\  \hline 
\end{tabular}
 \caption{Comparison of distributions based on Vuong's test.}
\label{table:comparisons}
\end{table}

There were 375 event times, ranging from 0.342 to 36 seconds. The jump in Figure~\ref{fig:cumulative_dist}B is an artifact of the coding: single punches and kicks were all coded as if they lasted one second, because in many cases, accurate time measurement of these behaviors was not feasible. This inaccuracy does not affect the maximum likelihood estimates; see $\tau_{min}$ in the bottom three rows in Table~\ref{table:stats}. 
In Figure~\ref{fig:cumulative_dist}B we can immediately see that the power law misfits the data. This is confirmed by Vuong's test, based on the tail of the lognormal ($\tau_{min} = 4.23; n = 93$). When comparing the lognormal to the power law, the lognormal is superior indeed (LL $= 3.46; p = 0.00027$), which confirms the second conjecture. To compare the exponential to the lognormal, the tail of the exponential is used ($\tau_{min} = 3.17; n = 108$). The exponential is not significantly better (LL $ = 1.03; p = 0.151$) and because it is visually indistinguishable from the lognormal, it is not depicted. 

\section*{Discussion}
In the scientific community, there is a broad consensus that interevent times are power-law distributed, based on coarse-grained data. As video data from CCTV camera's and mobile phones become more widely available, more fine-grained studies of violence become possible than in the past, which might challenge received wisdom. For the street fights that I examined, however, the power law was no clear winner but not refuted either. Perhaps the data was still too coarse-grained. After all, it should be (mathematically) possible for the interevent times to approximate zero duration \cite{sornette97}, which was not feasible when hand-coding the videos. This limitation might bias against the lognormal. However, there is also a bias against the power law, because bystanders lack motivation to keep filming long lulls. Consequently, the tail of the distribution is undersampled, which suggests that more CCTV data sets are necessary for a final verdict. Violent event times have no fat tailed distribution because fighting drains energy and therefore lasts shorter than the breaks in between. Therefore it should not surprise us that the lognormal distribution fitted better than the power law. The exponential's tail is shorter than the lognormal's tail but for this particular data, it did not fit better. Although the outcomes are somewhat inconclusive, representing (inter)event times as a multiplicative process yields a stronger argument than current theory where to expect power laws and where not. 

\subsection*{Data and Code}
The code and data are at \texttt{https://osf.io/bpsa6/} \\ The original coded video data is at \texttt{https://osf.io/f25nq/}

\footnotesize

\end{document}